\documentclass[aps,prl,groupedaddress,fleqn,twocolumn]{revtex4}
\usepackage{graphicx}
\usepackage{flushend}
\usepackage{balance}
\usepackage{physics}

\newcommand{\beq}{\begin{eqnarray}}
\newcommand{\eeq}{\end{eqnarray}}
\usepackage{amsmath}

\begin{document}

\title{Dynamical indistinguishability and statistics in quantum fluids}

\author{Alessio Zaccone$^{1,2}$}%
\author{Kostya Trachenko$^{3}$}%

\affiliation{$^{1}$Department of Physics ``A. Pontremoli'', University of Milan, via Celoria 16,
20133 Milan, Italy.}
\affiliation{$^{2}$Cavendish Laboratory, University of Cambridge, JJ Thomson
Avenue, CB30HE Cambridge, U.K.}
\affiliation{$^3$ School of Physics and Astronomy, Queen Mary University of London, Mile End, London, U.K.}

\begin{abstract}
For a system to qualify as a quantum fluid, quantum-statistical effects should operate in addition to quantum-mechanical ones. Here, we address the hitherto unexplored dynamical condition for the quantum-statistical effects to be manifested, and consider particle exchange events in the gaslike regime of fluid dynamics as a dynamical process with an intrinsic time scale. We subsequently propose a quantitative criterion of particle indistinguishability and associated quantum statistics to be inoperative at short time and emerge at long time. Verifiable experimentally, our predictions enable a systematic search for a transition between statistics-active and statistics-inactive regimes in quantum fluids.
\end{abstract}

\maketitle
Quantum statistics governs properties of quantum liquids and is related to different properties of 3He and 4He liquids \cite{pines,leggett1}. On the other hand, properties of solid $^3$He and $^4$He crystals are largely unaffected by quantum statistics and are mostly governed by conventional lattice-dynamics effects \cite{dobbs,leggett1}, suggesting that the ability of liquid particles to flow and exchange places is importantly related to particle indistinguishability and the emergence of quantum statistics. Another example is the observed difference in rotation and vibration spectra of heteronuclear and homonuclear molecules, underlying a difference between rotation where identical particles physically change places and vibration where they do not \cite{leggett1,leggett2}. Recognising the importance of the ability of particles to exchange places, Leggett formulates two criteria for a system to qualify as a quantum fluid \cite{leggett1,leggett2}. The first is the {\it quantum-mechanical} criterion: the wave nature of a particle should affect other particles. This is readily formulated as a condition that the de Broglie wavelength $\lambda_{\rm dB}$ should reach or exceed inter-particle separation $x$. The second criterion is {\it quantum-statistical}, stating that liquid particles are able to physically exchange places relatively easily in order to find out that they are indistinguishable and for the quantum statistics to emerge \cite{leggett2}.

Notably, this second dynamical criterion related to the ability of particles to exchange places has remained unexplored and unqualified - a surprising fact in view of the wide and long-standing interest in quantum fluids. It is therefore important to discuss physics involved in particle exchanges.

Indistinguishability in quantum mechanics is argued on the basis of impossibility to follow particle trajectories due to the Heisenberg uncertainty relation \cite{landau}. A related origin of indistinguishability is connected to the overlap of particle wave functions \cite{Bohm,Schrodinger}. Mathematically, indinstiguishability is expressed by writing the wave function of two particles as a symmetric or antisymmetric combination of one-particle wavefunctions, corresponding to Bose-Einstein and Fermi-Dirac statistics, respectively. For example, the latter case corresponds to the wave function

\begin{equation}
\psi=\frac{1}{\sqrt{2}}\left(f(x_1-x_a)g(x_2-x_b)-f(x_2-x_a)g(x_1-x_b)\right)
\label{psi}
\end{equation}
\noindent where $f(x_1-x_a)$ and $g(x_1-x_b)$ are the wave packets of the first particle and second particles in the regions of space near $x=x_a$ and $x=x_b$, respectively \cite{Bohm}.

Physically, symmetrisation \eqref{psi} means that no new quantum state is produced as a result of particle exchange where $x_1\leftrightarrow x_2$ \cite{Bohm}. The corresponding probability function $P$ is

\begin{eqnarray}
\begin{split}
& P=\frac{1}{2}\{|f(x_1-x_a)g(x_2-x_b)|^2+|f(x_2-x_a)g(x_1-x_b)|^2\\
& +[f^*(x_1-x_a)g(x_1-x_b)g^*(x_2-x_b)f(x_2-x_a)\\
& +{\rm complex~conjugate}] \}
\label{inter}
\end{split}
\end{eqnarray}

The first two terms represent ``classical probability'' terms, whereas the remaining terms are quantum-mechanical interference effects. The interference implies that particles can not be assigned a definite identity \cite{Bohm}.

We now observe that the ability of fluid particles to flow and exchange places and, consequently, exhibit effects of quantum statistics, is a dynamical process with a characteristic time scale \cite{frenkel}. Based on this observation, we develop a quantitative picture of the second Leggett criterion above in the gaslike regime of fluid dynamics. We find that at times shorter than relaxation time $\tau$, the particle exchanges are inoperative, and the statistics is classical. On the other hand, at times considerably exceeding $\tau$, the quantum statistics and ensuing properties set in. Verifiable experimentally, our predictions can be used to systematically search for a transition between statistics-active and statistics-inactive regimes.

It is convenient to first list the length scales in a fluid that are relevant to our discussion below. These are:

1. The mean free path $\ell=\frac{1}{n\sigma}$, where $n$ is concentration, $\sigma\propto a^2$ is particle cross-sectional area and $a$ is scattering length.

2. Average inter-particle separation $x=\frac{1}{n^\frac{1}{3}}$.

3. de Broglie wavelength $\lambda_{\rm dB}=\frac{h}{p}$, where $p$ is particle momentum.

4. Inter-particle separation in a condensed fluid state $d$ on the order of Angstroms. The order of magnitude of $d$ is set by fundamental physical constants as the Bohr radius $a_{\rm B}=\frac{\hbar^2}{m_ee^2}$, where $m_e$ and $e$ are electron mass and charge.

We start with considering a fluid consisting of quantum particles in the gaslike regime where particles move the mean free path distance $\ell$ before colliding and changing course. Swapping particle coordinates in the wave function \eqref{psi}, corresponding to the symmetrisation of the wave function, represents an exchange event between two particles, which we now consider as a dynamical process possessing an intrinsic time scale.

Particles moving in straight lines are unable to exchange places efficiently. In order for a physical exchange between two particles to take place, a particle needs to change its course and end up where another particle was before. Changing course takes place on the length scale given by $\ell$. This means that on average, pair exchange events are inoperative during time $t$ shorter than

\begin{equation}
\tau=\frac{\ell}{v}
\label{tau1}
\end{equation}
\noindent where $v$ is the average particle velocity.

If a system is dilute enough so that particle collisions are rare or absent, $\ell$ in Eq. \eqref{tau1} is substituted by another length, the system size, and $\tau$ can be substantial.

As mentioned earlier, quantum statistical mechanics provides a static or steady-state criterion for the observation of quantum effects in cold gases, $x=\lambda_{\rm dB}$ \cite{leggett1,leggett2}. When this condition is satisfied, quantum corrections to the gas equation of state become significant, and BEC itself becomes possible when $n(\lambda_{{\rm dB}})^{3}>2.612$ ~\cite{huang,bagnato}. This is an entirely ``static criterion'' for quantumness in the steady-state that the system eventually will reach, since it is derived from equilibrium quantum statistical mechanics, i.e. from consideration of partition functions in the steady state. The criterion \eqref{tau1} that we propose is instead based on a kinetic theory and on a different, dynamical, criterion for particle indistinguishability and quantum statistics to apply.

We therefore see that at times shorter than $\tau$, quantum indistinguishability and its consequences such as quantum statistics do not operate. As a result, the system exhibits Boltzmann statistics (for distinguishable quantum particles) on very short times, even if the steady-state condition $x=\lambda_{\rm dB}$ holds. On the other hand, particle exchanges in the fluid take place at longer times $t\gg\tau$. Together with the quantum-mechanical condition $x=\lambda_{\rm dB}$, this then results in quantum indistinguishability, and quantum statistics emerges at long times $t\gg\tau$.

We note that whilst Eq. \eqref{tau1} is related to a time scale of particle indistinguishability to manifest itself, there are other processes contributing to the experimental time lag. Non-equilibrium effects play a role in the kinetics of BEC formation \cite{zoran_2021}. Approaches based on quantum master kinetic equations~\cite{gardiner} show the ability to reproduce the kinetics of BEC formation and the initial lag time to a good accuracy~\cite{esslinger}. Other theories use Keldysh formalism to discuss non-equilibrium dynamics in the system where Bose statistics is already imposed \cite{stoof1,stoof2}. These effects are therefore related to non-equilibrium effects in the system where quantum statistics already set in and therefore are different to dynamical distinguishability discussed here.

$\tau$ increases as temperature decreases. If this temperature decrease is below the triple point pressure, Eq. \eqref{tau1} applies to all temperatures where the system is a gas. Above the pressure of the triple point, temperature decrease results in the gas-liquid phase transition. If pressure is higher still and above the critical point pressure, temperature decrease results in crossing the Frenkel line where, similarly to the gas-liquid transition, particle motion changes from gaslike diffusive motion above the line to the combined diffusive and oscillatory in the liquidlike state below the line \cite{ropp}. In both the liquid state and supercritical state below the Frenkel line, the concept of the mean free path $\ell$ does not apply. Instead, the new length scale appears: the inter-particle separation in a condensed state, $d$. $d$ is on the order of Angstroms in condensed matter phases, and is set by the overlap of outer electronic shells when chemical bonding forms. However, relaxation time can no longer be estimated as $\frac{\ell}{v}$ in the liquid or liquidlike states using Eq. \eqref{tau1}. Instead, the relevant relaxation time scale, known as liquid relaxation time $\tau_l$ \cite{frenkel}, acquires a different meaning and becomes the time between two consecutive particle jumps between adjacent quasi-equilibrium positions. $\tau_l$ is governed by an activated process and, differently from Eq. \eqref{tau1}, exponentially increases on lowering the temperature: $\tau_l\propto e^{\frac{U}{T}}$, where $U$ is the activation energy barrier \cite{frenkel}. $U$ is governed by liquid structure and interactions in a complicated way not amenable to analytical calculations (we note that at low temperature, tunneling can make a significant contribution to $\tau$, in which case $U$ sets the tunneling rate~\cite{weiss}). Hence, Eq. \eqref{tau1} applies to the gas state up to the boiling line temperature or gaslike state up to the Frenkel line temperature. Nevertheless, the criterion $t\ll\tau_l$, where classical statistics operates, still applies to the liquidlike dynamics regime, and implies that particle exchanges are inoperative at short times and, consequently, quantum indistinguishability and its implications such as quantum statistics do not affect experimental outcomes. The difference with the gaslike regime is that $\tau_l$ in the liquidlike regime can not be represented by $\tau$ in Eq. \eqref{tau1}.

$x$ is generally smaller than $\ell$ in Eq. \eqref{tau1}. If $x<\ell$, a more stringent criterion for the absence of particle exchanges is that the observation time should be shorter than

\begin{equation}
\tau_x=\frac{x}{v}
\label{taux}
\end{equation}

Using the quantum-mechanical criterion $x=\lambda_{\rm dB}$ in Eq. \eqref{taux}, gives $\tau_x=\frac{h}{mv^2}$, or

\begin{equation}
\tau_x=\frac{h}{k_{\rm B}T}
\label{tau3}
\end{equation}
\noindent where we used $v^2\approx\frac{k_{\rm B}T}{m}$.

Comparing $x$ to $\lambda_{\rm dB}$ is analogous to optics where the wave character of a ray becomes apparent when the size of the slit aperture is comparable to the wavelength, and where the wave behaves as a stream of particles if $\lambda$ is, instead, much smaller than the aperture slit. In our case, $x$ plays the role of the aperture slit, and $\lambda_{{\rm dB}}$ plays the role of the wavelength. We note that $\tau_x$ in Eq. \eqref{tau3} is the time it takes a particle to travel the distance equal to its de Broglie wavelength. Although $\frac{\lambda_{{\rm dB}}}{v}$ can be viewed as a single-particle property, the above discussion makes it relevant for quantum-statistical properties of a many-body system.

We recognise $\tau$ in Eq. \eqref{tau3} as the ``Planckian relaxation time'', $\tau_{\rm{Pl}}$ \cite{Zaanen-Nature,Zaanen:2018edk}. $\tau_{\rm{Pl}}$ was related to several fundamental physical phenomena, including the linear resistivity of strange metals \cite{Bruin804} and cuprate superconductors \cite{legros}, universal bounds on quantum chaos \cite{Maldacena:2015waa}, bounds on diffusion \cite{Hartnoll:2014lpa,Mousatov2020,Behnia_2019}, SYK model \cite{Patel:2019qce}, magic bilayer graphene \cite{2019arXiv190103710C}, black hole physics \cite{Sachdev:2015efa} and quark-gluon plasma \cite{qgp}. In these experiments, $\tau_{\rm{Pl}}$ emerges as a universal lower bound - a surprising result in view of different systems and processes involved. The relevance of $\tau_{\rm{Pl}}$ is often discussed on the basis of dimensionality arguments or energy-time/momentum-position uncertainty relations \cite{Zaanen-Nature,Zaanen:2018edk,Hartnoll:2014lpa,Mousatov2020,Bruin804} and is not fully understood \cite{Zaanen:2018edk}.
Here, $\tau_{\rm{Pl}}$ acquires the physical meaning as the time scale separating the regimes where particle indistinguishability and its consequences including the effects of quantum statistics set in. In future work it could be interesting to find a possible relation between $\tau_x$ above related to quantum indistinguishability and related interference terms in \eqref{inter} and the interpretation of $\tau_{\rm{Pl}}$ as the dissipation time in ``many-body entangled quantum matter'', argued using the insights from string theory and holography \cite{Zaanen:2018edk}.

We can estimate $\ell$ as $\ell=\frac{\sqrt{\pi}/8}{n\sigma}$, where $\sigma=8 \pi a^{2}$ is the total cross-section and $a$ is the scattering length \cite{huang}. For the system in ~\cite{BEC}, $a\approx 10^{-6}$~cm, while $n=2.6\times 10^{12}$~cm$^{-3}$, quoting from the original paper, yielding $\ell \approx 3.4 \times 10^{-3}$~cm, much larger than the inter-particle spacing $x$ estimated to be $7.3 \times 10^{-5}$~cm. Taking $v=\left(\frac{k_{\rm B}T}{m}\right)^{\frac{1}{2}}$ at the BEC temperature \cite{BEC} and using it in Eq. \eqref{tau1} gives $\tau$ of about $8$ ms. $\tau_x$ in \eqref{tau3} of about 0.3 ms at the experimental BEC temperature of 170 nanoKelvin \cite{BEC}. Both times are much shorter than the experimental lag time of BEC in the $^{87}$Rb to emerge on the order of seconds \cite{BEC}. This is useful to note because it shows that the criterion \eqref{tau1} could have been falsified had it implied a $\tau$ for the onset of quantum statistics longer than the experimental time scale at which BEC is seen, or longer than the thermalization time.

It is interesting to discuss whether our dynamical picture of quantum indistinguishability is experimentally verifiable. Let's assume that a system is prepared at time $t=t_0$ and observed until time $t_0+t$. If $t<\tau$ in \eqref{tau3}, our theory predicts that particle indistinguishability and quantum statistics does not affect observables. Instead, quantum statistics and other consequences of indistinguishability are predicted to set in at time $t\gg\tau$ only (this time may be well in excess of $\tau$ because $\tau$ signals the onset of particles exchanges only, whereas it may take considerably longer time for the macroscopic number of particles to undergo exchange events amounting to the symmetrization of the wave function).

One can systematically measure the waiting time of emergence of quantum-statistical effects in cold gases such as the condensate peak or Fermi surface \cite{fermi-surface} and compare it to the prediction in Eq. \eqref{tau1}. Doing so by varying temperature meets the challenge of changing the temperature faster than $\tau=\frac{\ell}{v}$ in \eqref{tau1} because $\frac{\ell}{v}$ is also a characteristic time during which the system thermalises and achieves thermal equilibrium. This implies that particle exchanges and their indistinguishability set in on the same time scale as thermalisation. This is nevertheless an important insight on its own: we can assume that a thermalised equilibrium system already attains the state of indistinguishability dynamically and hence fulfills the second Leggett criterion for the system to be a quantum fluid.

Another way to study this dynamical effect is to vary the particle scattering length $a$. $a$ can be varied very quickly using magnetic Feshbach resonance \cite{Hadzibabic}, with the associated quick change of $\ell$ in \eqref{tau1}. Reducing $a$ from its large value to small values below the BEC temperature results in the progressive increase of $\ell$ and $\tau$ and, according to above discussion, increasing waiting time for quantum-statistical effects, including BEC, to emerge. These effects can be studied as a function of waiting time for a range of $a$. In a recent experiment \cite{Hadzibabic}, a gas was quenched to large $a$, evolved for a range of waiting time $t_w$ and then quenched to small $a$ where the momentum distribution was measured. Here, we propose to reverse the last two steps: start with an equilibrated gas with large $a$, quench to a range of small $a$ and measure the momentum distribution and other quantum-statistical properties such as BEC as a function of the waiting time. We note that changing $a$ and $\ell$ may also induce a non-equilibrium state, however a relaxation of this state to equilibrium is of different nature as compared to that involving temperature change. It would also be interesting to compare the results of this experiment to that performed above the BEC temperature. This analysis can bring about interesting insights into the waiting times, their mechanisms and potential relation to the dynamical onset of indistinguishability and quantum statistics as well as widen our understanding of short-time non-equilibrium processes in fluids which are essential for understanding their thermodynamic and dynamical processes \cite{ropp}.

We note that the dynamics of particle exchanges is related to the degree of overlap of particle wave functions and associated energy terms such as the exchange energy. For a system of two electrons, for example, the transition rate $\Gamma$ between the state $|i\rangle=\phi_1(r_1)\phi_2(r_2)$ and the state with particles swapped, $|f\rangle=\phi_1(r_2)\phi_2(r_1)$, is $\Gamma\propto|\langle f|U|i\rangle|^2$ according to the Fermi golden rule, where $U$ is the perturbation interaction operator and $\langle f|U|i\rangle$ is its matrix element

\begin{equation}
\langle f|U|i\rangle=\int U\phi_1(r_1)\phi_1^*(r_2)\phi_2(r_2)\phi_2^*(r_1)dV_1dV_2
\label{matrix}
\end{equation}

We recognize \eqref{matrix} as the exchange energy $J$ setting the energy levels of two electrons and depending on the overlap of the wave functions $\phi_1(r_1)$ and $\phi_2(r_2)$ \cite{landau}. In this simple model, the transition rate and the associated exchange energy term $J$ are related as $\Gamma\propto J^2$. $\Gamma$ is related to $\tau=\frac{\ell}{v}$ in \eqref{tau1} in a sense that both depend on $U$ because the cross-sectional area $\sigma$ in $\ell=\frac{1}{n\sigma}$ is governed by $U$. $\tau$ additionally depends on other parameters such as temperature and concentration.

Theoretically, one significance of this picture lies in viewing the symmetrisation such as \eqref{psi} and \eqref{inter} not as a virtual event as considered previously, but as a physical exchange of particle wave packets with an associated dynamical process and time scale. This implies that, for example, interference and correlations in the last two terms in (2) can not propagate faster than the speed set by particle exchanges. This bears relation to the discussions of entanglement issues that have been of long-standing interest such as causality and locality. This relation can be developed in future work.

In summary, we formulated a quantitative criterion for quantum indistinguishability and ensuing quantum-statistical effects to become inoperative at short time and emerge in the long-time regime. Verifiable experimentally, our predictions enable a systematic search for a transition between statistics-active and statistics-inactive regimes.

We are grateful to E. Artacho, K. Behnia, V. Brazhkin, Z. Hadzibabic and J. Zaanen for discussions and EPSRC for support.

\bibliographystyle{apsrev4-1}

\bibliography{sample}

\end{document}